\documentclass[a4paper,11pt]{article}
\pdfoutput=1
\usepackage{jcappub}

\usepackage{color}
\usepackage{amsfonts}
\usepackage{graphicx}
\usepackage{amssymb}

\usepackage{eufrak}
\usepackage{mathrsfs}

\usepackage{mathtools}
\usepackage{ulem}
\usepackage{bm}
\usepackage{xcolor}
\renewcommand{\thesection}{\arabic{section}}

\def\laq{~\raise 0.4ex\hbox{$<$}\kern -0.8em\lower 0.62ex\hbox{$\sim$}~}
\def\gaq{~\raise 0.4ex\hbox{$>$}\kern -0.7em\lower 0.62ex\hbox{$\sim$}~}

\def\beq{\begin{equation}}
\def\eeq{\end{equation}}
\def\bea{\begin{eqnarray}}
\def\eea{\end{eqnarray}}

\def \ra {\rightarrow}
\def \ti {\widetilde}

\def \Mp {M_{\rm P}}

\def \da {\delta}
\def \b {\beta}
\def \a {\alpha}
\def \ap {\alpha^{\prime}}

\def \sg {\sigma}

\def \da {\delta}
\def \ep {\epsilon}
\def \r {\rho}
\def \om {\omega}
\def \Om {\Omega}

\def \car {{\cal R}}

\title{Primordial Black Holes\\ from Pre-Big Bang inflation}

\author[a]{P. Conzinu,}
\author[b]{M. Gasperini,}
\author[a]{and G. Marozzi}

\affiliation[a]{Dipartimento di Fisica, Universit\`a di Pisa, Largo B. Pontecorvo 3, 56127 Pisa, 
Italy,\\
and Istituto Nazionale di Fisica Nucleare, Sezione di Pisa, Italy}
\affiliation[b]{
Dipartimento di Fisica, Universit\`a di Bari, 
Via G. Amendola 173, 70126 Bari, Italy,\\
and Istituto Nazionale di Fisica Nucleare, Sezione di Bari, Italy
}

\emailAdd{pietro.conzinu@virgilio.it}
\emailAdd{gasperini@ba.infn.it}
\emailAdd{giovanni.marozzi@unipi.it}

\abstract{We discuss the possibility of producing a significant fraction of dark matter in the form of primordial black holes in the context of the pre-big bang inflationary scenario. We take into account, to this purpose, the enhancement of curvature perturbations possibly induced by a variation of the sound-speed parameter $c_s$ during the string phase of high-curvature inflation. After imposing all relevant observational constraints, we find that the considered class of models is compatible with the production of a large amount of primordial black holes in the mass range relevant to dark matter, provided the sound-speed parameter is confined in a rather narrow range of values, $0.003 \laq c_s \laq 0.01$. }

\keywords{primordial black holes, 
cosmological perturbation theory,
string cosmology, inflation
 
\vskip13pt plus8pt minus11pt

\noindent{\bfseries\large\sffamily{Preprints:}} BA-TH/800-20
}

\begin{document}

\maketitle

%%%%%%%%%%%%%%%%%%%%%%%%%%%%%%%%%%%%%%%%%%%%%%%%%%%%%%%%%%%%%%%%%%%%%%%%%%%

\section{Introduction}
\label{Sec1}
\setcounter{equation}{0}

The possibility that relatively light, but not yet evaporated, primordially produced black holes may contribute to a non-negligible fraction of the total dark matter density has been suggested long ago \cite{1,2,3}, and has recently been the subject of many interesting studies (see e.g. \cite{4,5,5a,5b,6,7,8,8a,9,10} and references therein).

Existing astrophysical/cosmological constraints \cite{10a,10b} seem to leave open, for such Primordial Black Holes (PBHs), a rather narrow mass window \cite{11,12,12a}: $M \sim 10^{18} - 10^{22}$g. In that mass range PBHs can be formed by the collapse of primordial inhomogeneities re-entering the horizon, and may give a significant contribution to the present dark matter density provided the associated curvature perturbations (sourcing the matter overdensity) have a primordial spectral amplitude $P(\om_M) \gaq 10^{-2}$, where $\om_M$ are the frequency modes crossing the horizon at the epoch of PBH formation (see e.g. \cite{13,14,15} for recent reviews, and \cite{15a,15b} for previous work on PBH formation). 

The aim of this paper is to show that, in models of pre-big bang inflation \cite{16,17} with a sufficiently long string phase of constant, high curvature and linearly growing dilaton, it is possible to obtain an allowed region of parameter space compatible with an efficient production of PBH-dark matter, provided, during the high-curvature string phase, the propagation of the background perturbations is characterized by an appropriate value of the sound speed parameter $c_s$ smaller than one. As we shall see, however, the production is efficient only in a very small range of values for $c_s$.

It should be recalled, at this point, that the possible enhancement of curvature  perturbations due to a sound speed $c_s<1$, in the spectral range relevant to PBH production, is not  a new effect but a possibility already discussed in the literature \cite{17a,18,19,28a} (see also \cite{19a}). The physical origin of the ``modified" effective velocity ($c_s \not=1$) considered here, however, is different from the one of  previous papers. The resulting value $c_s \not= 1$ is not due to particular properties of the effective potential controlling the evolution of the inflaton field, or to a non-standard kinetic term like in the context of Dirac-Born-Infeld (DBI) inflation, as discussed in \cite{17a}. The modified $c_s$ is possibly due, instead, to the background $\ap$ corrections unavoidably predicted by string theory, and producing an effective sound-speed parameter different from one (and constant) for the whole duration of the high-curvature string phase preceding the transition to the standard decelerated evolution (as shown long ago for the case of tensor metric perturbations \cite{20}). The effective value of $c_s$ depends, of course, on the specific form of the given $\ap$ corrections, and it would be interesting, in this respect, to evaluate $c_s$ in the context of the non-perturbative backgrounds solving the string cosmology equations to all orders in $\ap$ (see e.g. \cite{43,44,45}).

Let us also stress that the contribution of the $\ap$ corrections may produce, in general, an effective value $c_s \not=1$ (and a consequent modification of the spectral amplitude) for all types of background fluctuations. In our context, in particular, for the tensor perturbations of the metric background, for the scalar perturbations (sourced by the dilaton), and for the perturbations of the Kalb-Ramond axion field, associated by space-time duality with the four-dimensional components of the NS-NS (Neveu-Schwarz) two-form appearing in the string effective action (see e.g. \cite{21}). The presence of this last type of fluctuations, in particular, is crucial for the production -- via the curvaton mechanism \cite{22} -- of a viable spectrum of adiabatic curvature perturbations \cite{23,24}.

However, the modification of the spectrum of tensor perturbations due to a modified sound velocity has no impact to first order  on PBH production, which directly depends only on scalar-type inhomogeneities (it may have some role, however, in restricting the allowed values of the model parameters through existing phenomenological constraints, see \cite{25,26}). In addition, we have explicitly checked with preliminary numerical computations  (see also \cite{tesi}) that the spectral distribution of the scalar metric and dilaton perturbations, directly amplified by the phase of pre-big bang inflation (and unavoidably characterized by a strongly tilted ``blue" spectrum \cite{27,28}), is so steep, and thus so depressed outside the band of very high frequency modes, to leave very little room in parameter space for a significant PBH production (even if the sound-speed parameter is quite different from one).

The primordial spectrum of the axionic fluctuations, on the contrary, turns out to be amplified (by the same dynamics) with a much ``flatter" distribution \cite{29,30}, and is thus much more sensitive to variations of the speed of sound, and to a possibly related PBH production. In this paper, where we take an illustrative and phenomenological approach, we will thus take into account a possible variation of the speed of sound for axion perturbations only, keeping $c_s=1$, for simplicity, for both scalar and tensor metric perturbations. But let us anticipate that -- at least for the class of models we are considering --  the results are qualitatively the same as those we would obtain by including a possible variation of $c_s$ for all types of perturbations. This has recently been confirmed in \cite{tesi}, by explicitly taking into account also the modified sound velocity of scalar metric and dilaton perturbations. Let us finally stress that, motivated by the example discussed in \cite{20}, and by the practically ``step-like" variation of the sound velocity induced in that case by the $\ap$ corrections, we shall adopt here a sudden approximation to describe the variation of the sound velocity from the value $c_s=1$ to the constant value typical of the string phase.

It should be stressed that the possibility of PBH production in the context of the pre-big bang scenario was first pointed out and discussed in \cite{45a}. 
The work presented in this paper is also motivated by the natural expectation that a high-curvature phase of pre-big bang inflation, possibly dominated by a gas of ``stringy" black holes (as shown in \cite{42}), and characterized by a blue spectrum of scalar metric perturbations, should also lead to an abundant formation of PBH-dark matter \cite{32a}. This was also the spirit of previous papers studying PBH formation due to a cosmic contracting phase preceding the bounce (see in particular \cite{47a,47b,47c}).
Despite such an expectation, it turns out  -- for the class of models considered in this paper -- that no significant PBH production is possible at all, via the standard collapse of primordial inhomogeneities, if we use the predicted spectrum of perturbations \cite{23,24,27,28} neglecting the possibility of $c_s \not=1$, and/or we try to obtain an enhanced spectral amplitude for the relevant frequency modes by varying other spectral parameters without violating present limits on non-Gaussianity.  The only mechanism which seems to work, consistently with all present observational constraints, is the one discussed in this paper. 

The paper is organized as follows. In Sect. \ref{Sec2} we compute the primordial spectrum of scalar perturbations generically associated to a model of inflation with two different phases of accelerated evolution, one with sound-speed parameter $c_s=1$ and the other characterized by a constant value $c_s<1$. In Sect. \ref{Sec3} we explicitly introduce our class of models of pre-big bang inflation, their parameters, and the associated spectra of scalar and tensor perturbations. In Sect. \ref{Sec4} we impose on the parameters all relevant constraints required to satisfy self-consistency conditions, as well as to match the produced spectra with present observational data. In Sect. \ref{Sec5} we show that, for appropriate values of the sound velocity, we are left with an allowed region in parameter space compatible with the production of a large fraction of PBH dark matter. Finally, in Sect. \ref{Sec6} we present our final remarks and conclusions. 

%%%%%%%%%%%%%%%%%%%%%%%%%%%%%%%%%%%%%%%%%%%%%%%%%%%%%%%%%%%%%%%%%%%%

%%%%%%%%%%%%%%%%%%%%%%%%%%%%%%%%%%%%%%%%%%%%%
\section{Two phases of inflation with different values of the sound-speed parameter}
\label{Sec2}
\setcounter{equation}{0}
%%%%%%%%%%%%%%%%%%%%%%%%%%%%%%%%%%%%%%%%%%%%%%%%

Let us consider a simple, but generic, model of inflationary background described by a spatially flat FLRW metric, and including a first phase of accelerated evolution extending in conformal time from $\eta =-\infty$ to $\eta=-\eta_s$, and a second one (with possibly different kinematics) from $\eta =-\eta_s$ to $\eta=-\eta_1<0$. For $\eta>-\eta_1$ the background enters the regime of standard decelerated expansion. We have in mind, in particular, a string cosmology scenario with an initial, low-energy phase of dilaton-driven inflation, followed by a string phase of high-curvature inflation (see Sect. \ref{Sec3}), but, in this Section, we will keep our discussion as general as possible. 

Let us consider in this background a scalar fluctuation $\car$, described by a canonically normalized variable $v= z \car$ which diagonalizes the kinetic part of the perturbed action, and satisfies the canonical equation (see e.g. \cite{31})
\beq
v'' - \left( c_s^2 \nabla^2 + {z''\over z} \right) v=0\,.
\label{21}
\eeq
Here a prime denotes differentiation with respect to the conformal time $\eta$, $z(\eta)$ is the so-called ``pump field" controlling the evolution (and amplification) of the given perturbations, and the parameter $c_s$ plays the role of an effective sound-velocity coefficient. Both $z$ and $c_s$ are fully determined by the given background fields and sources.

We shall assume for the pump field a power-law behavior, $z \sim |\eta|^\a$, and we shall work in the so-called ``sudden approximation" \cite{32} in which the background undergoes a sudden variation of the pump field and a step-like variation of the sound-speed parameter. We will suppose, in particular, that the sound velocity has a non-trivial (but constant) value only during the second inflationary phase, i.e.
\beq
c_s = {\rm const} \not= 1 ~~{\rm if}~~  -\eta_s < \eta < -\eta_1
~~~~~~,~~~~~~~
c_s = 1 ~~{\rm if}~~ \eta<-\eta_s ~~{\rm or}~~ \eta >-\eta_1,
\label{22}
\eeq
and that the (continuous) canonically normalized pump field is given by
\bea
z_1&=&{\Mp\over \sqrt{2}} \left(\eta_s\over \eta_1\right)^{\a_2} \left(-\eta\over \eta_s\right)^{\a_1}, ~~~~~~~~~~~~~~~~~~~~ \eta < -\eta_s,
\nonumber \\ 
z_2&=&{\Mp\over \sqrt{2}} \left(-\eta\over \eta_1\right)^{\a_2}, ~~~~~~~~~~~~~~~~~~~~~~~~~~~ -\eta_s< \eta < -\eta_1,
\nonumber \\ 
z_3&=&{\Mp\over \sqrt{2}} \left(\eta+2\eta_1\over \eta_1\right)^{\a_3}, ~~~~~~~~~~~~~~~~~~~~~ -\eta_1 < \eta,
\label{23}
\eea
where $\Mp$ is the Planck mass and $\a_1$, $\a_2$ and $\a_3$ are constants depending on the background dynamics. The canonical equation for the Fourier modes $v_k$ (such that $\nabla^2 v_k =-k^2 v_k$) thus takes the form
\bea
&&
v_k''+\left[k^2 -{\a_1(\a_1-1)\over \eta^2}\right] v_k=0, ~~~~~~~~~~~~~~~~~~~~~~~~~ \eta < -\eta_s,
\nonumber \\ 
&&
v_k''+\left[c_s^2k^2 -{\a_2(\a_2-1)\over \eta^2}\right] v_k=0,~~~~~~~~~~~~~~~~~~ -\eta_s< \eta < -\eta_1,
\nonumber \\ 
&&
v_k''+\left[k^2 -{\a_3(\a_3-1)\over (\eta+2\eta_1)^2}\right] v_k=0,
~~~~~~~~~~~~~~~~~~~~~ -\eta_1 < \eta.
\label{24}
\eea
It may be noted that, in the third phase describing the decelerated evolution of the standard cosmological scenario, the pump field $z_3$ for metric perturbations is expected to coincide with the scale factor, i.e. $z_3(\eta) \sim \Mp \,\eta^{\a_3} \sim \Mp\, a(\eta)$. In a string cosmology context, and during the phase of pre-big inflation, the pump field may depend instead on the time evolution of both the scale  factor and the dilaton, even in the simple case of scalar metric perturbations (see Sect. \ref{Sec3}). 

The general solution of the mode equations (\ref{24}) can now be written in terms of the first-kind and second-kind Hankel functions as \cite{33}
\bea
&&
v_k^1= \left(\pi \eta\over 4\right)^{1/2} \left[c_+^1 H_{\nu_1}^{(2)}(k\eta) +
c_-^1 H_{\nu_1}^{(1)}(k\eta)\right],
 ~~~~~~~~~~~~~~~~~~~~~~~~~~~~~~~~~~~~ \eta < -\eta_s,
\nonumber \\ 
&&
v_k^2= \left(\pi \eta\over 4\right)^{1/2} \left[c_+^2 H_{\nu_2}^{(2)}(c_s k\eta) +
c_-^2 H_{\nu_2}^{(1)}(c_s k\eta)\right],
~~~~~~~~~~~~~~~~~~~~~~~~~~~~ -\eta_s< \eta < -\eta_1,
\nonumber \\ 
&&
v_k^3= \left(\pi (\eta+2\eta_1)\over 4\right)^{1/2} \left[c_+^3 H_{\nu_3}^{(2)}(k\eta+2 k\eta_1) +
c_-^3 H_{\nu_3}^{(1)}(k\eta+2k\eta_1)\right],
~~~ -\eta_1 < \eta,
\label{25}
\eea
where $c_{\pm}^{1,2,3}$ are integration constant, and where the Bessel indices $\nu_i$ are determined by the pump field kinematics as $\nu_i= 1/2-\a_i$, with $i=1,2,3$. Let us impose on the canonical modes the initial normalization (for $\eta \ra -\infty$) to the quantum fluctuations of the vacuum, given by (the adiabatic vacuum)
\beq
v_k ={1\over \sqrt{2k}} e^{-ik\eta +i\phi}, ~~~~~~~~~~~~~~~~~~~~~~~~~~
\eta \ra -\infty,
\label{26}
\eeq
where $\phi$ is an arbitrary initial phase. By using the large argument limit of the Hankel functions \cite{33},
\beq
 H_{\nu}^{(2)}(x) =\sqrt{2\over \pi x} e^{-ix-i\ep_\nu}, ~~~~~~~~~~
 H_{\nu}^{(1)}(x) =\sqrt{2\over \pi x} e^{ix+i\ep_\nu}, ~~~~~~~~ x \ra \infty,
 \label{27}
 \eeq
 where $\ep_\nu= -(\pi/2)(\nu+1/2)$, we obtain $c_+^1=1$, $c_-^1=0$, and we can write the normalized solution for the scalar perturbation $\car_k=v_k/z$ as follows:
 
 \bea
&&
 \!\!
\car_k^1= {v_k^1\over z_1}={1\over \Mp}\left(-\pi \eta_s\over 2\right)^{1/2} 
 \left(\eta_s\over \eta_1\right)^{-\a_2} \left(-\eta\over \eta_s\right)^{\nu_1}
H_{\nu_1}^{(2)}(k\eta),
 ~~~~~~~~~~~~~~~~~~~~~~~~~~ \eta < -\eta_s,
\nonumber \\ 
&&
 \!\!
\car_k^2={v_k^2\over z_2}={1\over \Mp}\left(-\pi \eta_1\over 2\right)^{1/2} 
 \left(-\eta\over \eta_1\right)^{\nu_2}
\left[c_+^2 H_{\nu_2}^{(2)}(c_s k\eta) +
c_-^2 H_{\nu_2}^{(1)}(c_s k\eta)\right],
~~~~ -\eta_s< \eta < -\eta_1,
\nonumber \\ 
&&
 \!\!
\car_k^3={v_k^3\over z_3}={1\over \Mp}\left(\pi \eta_1\over 2\right)^{{1\over 2}}
\left(\eta+2\eta_1\over \eta_1\right)^{\nu_3}
\left[c_+^3 H_{\nu_3}^{(2)}(k\eta+2 k\eta_1) +
c_-^3 H_{\nu_3}^{(1)}(k\eta+2k\eta_1)\right],
 -\eta_1 < \eta
 \nonumber \\ 
&&
\label{28}
\eea
(we have chosen the initial phase $\phi=-\ep_{\nu_1}$). It should be noted that the sound-speed $c_s$ explicitly appears only in the argument of the solutions $\car_k^2$, but it is also implicitly contained inside the constants $c_{\pm}^{2,3}$ (see below). Note also that, with the chosen normalization, our integration constants $c_{\pm}^{2,3}$ may represent, in a second-quantization context, the so-called Bogoliubov coefficients controlling the expectation value of the number of scalar ``quanta" produced out of the vacuum thanks to the amplification on the initial fluctuations (see e.g. \cite{21}). 

We are now interested in the primordial amplitude of the perturbation modes when re-entering the horizon, after the two phases of accelerated evolution. More precisely, we want to evaluate the dimensionless power spectrum
\beq
P(k) = {k^3\over 2 \pi^2} \left| \car_k^3\right|^2_{|k\eta|=1},
\label{29}
\eeq
for $\eta > -\eta_1$.
This is the relevant quantity controlling the collapse probability of the primordial inhomogeneities, and the associated possible formation of PBHs.  To this purpose we have to compute the coefficients $c_{\pm}^{2,3}$ by imposing the appropriate matching conditions on the modes $\car$ at the two transition epochs $-\eta_s$ and $-\eta_1$. 

It should be stressed, in this context, that the requirement of continuity of the background fluctuations $\car$ is not equivalent in general to the continuity of the canonical variable $v$, if the matching conditions are imposed on a discontinuous background \cite{34}. The continuity of $\car$, on the other hand, is needed to guarantee the continuity of the total energy density across the matching hypersurface \cite{35}. 

Let us then start by imposing the continuity of $\car$ and of its first derivative $\car'$ at $\eta=-\eta_s$, namely
\beq
\car_k^1(-\eta_s) = \car_k^2(-\eta_s), ~~~~~~~~~~~~~~~
\car_k^{\prime 1}(-\eta_s) = \car_k^{\prime 2}(-\eta_s),
\label{210}
\eeq
which gives us an inhomogeneous system of two algebraic equations for the two unknown variables $c_{\pm}^2$. It may be convenient, in this context, to introduce the notation $k\eta_s \equiv x_s$, and to use the following properties of the Hankel functions and of the derivatives with respect to their argument:
\beq
H_{\nu}^{(2)}(-x)=-e^{i\pi\nu} H_{\nu}^{(1)}(x), ~~~~~~~~~~~
{dH_{\nu}^{(2)}\over dx} (-x)= e^{i\pi\nu} {dH_{\nu}^{(1)}\over dx} (x).
\label{211}
\eeq
We then find that the determinant of the coefficients of our algebraic system can be simplified by using the following Wronskian condition \cite{33}:
\beq
H_{\nu_2}^{(1)}(x){d\over dx} H_{\nu_2}^{(2)}(x)- 
H_{\nu_2}^{(2)}(x){d\over dx} H_{\nu_2}^{(1)}(x)= -{4i\over \pi x}.
\label{212}
\eeq
Also, we can eventually eliminate the derivatives of the Hankel functions through the  relation
\beq
{d\over dx} H_{\nu}^{(1,2)}(x)={\nu\over x} H_\nu^{(1,2)} (x) -  H_{\nu+1}^{(1,2)} (x), 
\label{213}
\eeq
and write the final result as:
\bea
e^{i\pi\nu_2}c_+^2&=&{i\pi\over 4} c_s x_se^{i\pi\nu_1} \left[ {1\over c_s}H_{\nu_1+1}^{(1)} (x_s)H_{\nu_2}^{(2)} (c_sx_s)+2{\nu_2-\nu_1\over c_s x_s} H_{\nu_1}^{(1)} (x_s)H_{\nu_2}^{(2)}(c_sx_s)- \right.
\nonumber \\ && \left. ~~~~~~~~~~~~~~~~~~
-H_{\nu_1}^{(1)} (x_s)H_{\nu_2+1}^{(2)}(c_sx_s)
\right],
\label{214}\\ 
e^{-i\pi\nu_2}c_-^2&=&{i\pi\over 4} c_s x_se^{i\pi\nu_1} \left[ -{1\over c_s}H_{\nu_1+1}^{(1)} (x_s)H_{\nu_2}^{(1)} (c_sx_s)+2{\nu_1-\nu_2\over c_s x_s} H_{\nu_1}^{(1)} (x_s)H_{\nu_2}^{(1)}(c_sx_s)+ \right.
\nonumber \\ && \left. ~~~~~~~~~~~~~~~~~~
+H_{\nu_1}^{(1)} (x_s)H_{\nu_2+1}^{(1)}(c_sx_s)
\right]\,.
\label{215}
\eea

A similar computation, starting with the continuity condition for our  perturbation variable at the second transition epoch $\eta =-\eta_1$,
\beq
\car_k^2(-\eta_1) = \car_k^3(-\eta_1), ~~~~~~~~~~~~~~~
\car_k^{\prime 2}(-\eta_1) = \car_k^{\prime 3}(-\eta_1),
\label{216}
\eeq
and using the notation $k\eta_1 \equiv x_1$, leads to the result
\bea
&&
c_+^3=-{i\pi\over 4} x_1 \left\{
\left[c_+^2e^{i\pi\nu_2}H_{\nu_2}^{(1)}(c_s x_1) +
c_-^2 e^{-i\pi\nu_2} H_{\nu_2}^{(2)}(c_s x_1)
\right]
\left[2{\nu_2+\nu_3\over x_1}  H_{\nu_3}^{(1)}( x_1)-  H_{\nu_3+1}^{(1)}( x_1)\right]
\right.
\nonumber \\ && 
\left. ~~~~~~~~~~~~~~~~~~~
-c_s  H_{\nu_3}^{(1)}( x_1)\left[c_+^2e^{i\pi\nu_2}H_{\nu_2+1}^{(1)}(c_s x_1) +
c_-^2 e^{-i\pi\nu_2} H_{\nu_2+1}^{(2)}(c_s x_1)
\right] \right\},
\label{217}
\eea
\bea
&&
c_-^3={i\pi\over 4} x_1 \left\{
\left[c_+^2e^{i\pi\nu_2}H_{\nu_2}^{(1)}(c_s x_1) +
c_-^2 e^{-i\pi\nu_2} H_{\nu_2}^{(2)}(c_s x_1)
\right]
\left[2{\nu_2+\nu_3\over x_1}  H_{\nu_3}^{(2)}( x_1)-  H_{\nu_3+1}^{(2)}( x_1)\right]
\right.
\nonumber \\ && 
\left. ~~~~~~~~~~~~~~~~~~~
-c_s  H_{\nu_3}^{(2)}( x_1)\left[c_+^2e^{i\pi\nu_2}H_{\nu_2+1}^{(1)}(c_s x_1) +
c_-^2 e^{-i\pi\nu_2} H_{\nu_2+1}^{(2)}(c_s x_1)
\right] \right\}.
\label{218}
\eea

We are now in the position of evaluating the power spectrum (\ref{29}). Let us note, for an approximate evaluation (which is enough to our purposes), that there are in principle four different branches of the spectrum depending on the numerical value of the Hankel arguments $c_sx_s$, $x_s$, $x_1$, $c_sx_1$, which may be larger or smaller than one thus introducing, in the computations of $c_{\pm}^{2,3}$, the large- or the small-argument regime of the corresponding Hankel functions. We shall  consider now the various possible cases, by concentrating our discussion, in particular, on a (possibly realistic) scenario with $c_s<1$. It may be useful to define, in such a context, the frequency scales $k_s=1/\eta_s$ and $k_1=1/\eta_1$, so that we can also write $ x_s \equiv k/k_s$ and $x_1 \equiv k/k_1$. Of course, $\eta_s >\eta_1$ and $k_s<k_1$. 

Let us start with the highest frequency modes in the band defined by $k_1<k<k_1/c_s$, and thus characterized by the conditions:
\beq
x_s>1,~~~~~~~ c_s x_s>1, ~~~~~~ x_1>1,~~~~~~ c_sx_1<1.
\label{219}
\eeq
We note that according to Eq. (\ref{24}) the mode with $k \sim (c_s \eta_1) = k_1/c_s$ is indeed the last (i.e. highest) frequency mode to exit the horizon and to be amplified by inflation. Higher modes (which do not ``hit" the effective potential barrier present in the canonical equation, due to the inflationary pump-field evolution) are to be eliminated by a sharp UV cutoff, which has to be imposed on the spectrum because, for such modes, the adopted sudden approximation is not appropriate. By replacing the step transition with a smooth one, we would find indeed that the spectral amplitude of such modes with $c_sk\eta_1>1$ turns out to be exponentially suppressed as $\exp(-c_s k\eta_1)$, thus avoiding UV divergences \cite{36,37}. But for the purpose of this paper it will be enough to simply drop such an ultra-high freqeuncy tail of the spectrum, limiting ourselves to modes with $k<k_1/c_s$. 

For the frequency band specified by Eq. (\ref{219}), corresponding to modes crossing the horizon when $c_s<1$  and immediately re-entering at the end of the string phase (because of the sudden change of $c_s$), 
the coefficients $c_{\pm}^2$, defined in Eqs. (\ref{214}) and (\ref{215}) can be approximated as follows:
\bea
e^{i\pi\nu_2}c_+^2&=&{1\over 2} e^{i{\pi\over 2}(\nu_1+\nu_2)} e^{i(1-c_s)x_s} \left(\sqrt{c_s} + {1\over \sqrt{c_s}}\right),
\nonumber \\ 
e^{-i\pi\nu_2}c_-^2&=&-{i\over 2} e^{i{\pi\over 2}(\nu_1-\nu_2)} e^{i(1+c_s)x_s} \left(\sqrt{c_s} - {1\over \sqrt{c_s}}\right)
\label{220}
\eea
(we have applied the large argument limit (\ref{27}) to the Hankel functions of Eqs. (\ref{214}) and (\ref{215})). It can be easily checked that, for $c_s=1$, one recovers the standard result $|c_+^2|^2=1$, $|c_-^2|^2=0$, which means that such high frequency modes are unaffected by the first background transition -- namely, they are still well inside the horizon at the epoch $\eta=-\eta_s$, and will cross the horizon only during the second phase of inflation. In our case, however, their amplitude is affected by the variation of the sound-speed parameter, and we can write our approximate result (for $c_s<1$) as 
\beq
e^{\pm i\pi\nu_2}\,c_\pm^2= \a_\pm \,c_s^{-1/2},
\label{221}
\eeq
with $|\a_\pm| $ a number of order one.

For the analogous computation of $c_\pm^3$ we start with the exact Eqs. (\ref{216}) and (\ref{217}) and -- as prescribed by Eq. (\ref{219}) -- we take the large-argument limit of $H_{\nu_3}^{(1,2)}( x_1)$ but, for $H_{\nu_2}^{(1,2)}(c_s x_1)$, we have to use  the small argument limit, which is given in general by \cite{33}:
\beq
H_{\nu}^{(1)}(x) = p_\nu x^\nu+ i q_\nu x^{-\nu} + \cdots, ~~~~~~
H_{\nu}^{(2)}(x) = p_\nu^* x^\nu- i q_\nu x^{-\nu} + \cdots. 
\label{222}
\eeq
The coefficients $q$ and $p$ are dimensionless numbers with modulus of order one (when $\nu=0$ there are additional logarithmic corrections, i.e. $H_0^{(1)} \sim p_0 +iq_0 \ln x + \cdots$). A simple, explicit computation then gives, to leading order, 
\beq
\left|c_\pm^3\right| \sim x_1 (c_s x_1)^{-|\nu_2|-1/2}~,
\label{223}
\eeq
modulo a numerical coefficient of order one. Where we have used Eq. (\ref{221}) for $c_\pm^2$, and we have also taken into account the possibility that $\nu_2 <0$.

The power spectrum (\ref{29}) can then be estimated by recalling the explicit form of the solution $\car_k^3$ given in Eq.(\ref{28}). Using again the large-argument limit for $x_1$ we find, to leading order\footnote{We have applied the standard definition (2.9) for the spectral amplitude at re-entry, but let us stress again, for this frequency band,  that all modes are characterised by an almost instantaneous re-enter inside the horizon just at the end of the string phase, because of the sudden change of $c_s$. Actually, the result (\ref{224})  keeps unchanged if $\car_k^3$ is evaluated at  $\eta=-\eta_1$.},
\beq
\left|\car_k^3\right|_{|k\eta|\sim 1} \sim {1\over \Mp} {\left|c_\pm^3\right|\over \sqrt k}~,
\label{224}
\eeq
from which
\beq
P(k) \sim k^3 \left|\car_k^3\right|^2_{|k\eta|\sim 1} 
\sim \left(k_1\over \Mp\right)^2 
\left(k\over k_1\right)^{3-2|\nu_2|} c_s^{-1-2|\nu_2|}~,
~~~~~~~~~ k_1<k<k_1/c_s.
\label{225}
\eeq
As a final remark we may recall that at the epoch $\eta=-\eta_1$, which marks the end of inflation and the beginning of the standard cosmological scenario, the pump field $z_3$ should reduce (as previously stressed) to the scale factor, i.e. $z_3=\Mp a/\sqrt{2}$. Comparing with Eq. (\ref{23}) we get $a_1 \equiv a(-\eta_1)=1$, so that the frequency scale $k_1$ can be directly expressed in terms of the background curvature scale at the epoch $-\eta_1$, i.e. the (horizon crossing) Hubble scale $H_1\equiv H(-\eta_1)$ such that $k_1 \sim H_1 a_1$. Hence, our previous spectrum can be more conveniently parametrized also as follows:
\beq
P(k) \sim k^3 \left|\car_k^3\right|^2_{|k\eta|\sim 1} 
\sim \left(H_1\over \Mp\right)^2 
\left(k\over k_1\right)^{3-2|\nu_2|} c_s^{-1-2|\nu_2|}~,
~~~~~~~~~ k_1<k<k_1/c_s.
\label{226}
\eeq
Note that there is no dependence at all on the power $\nu_3$, i.e. on the kinematics of the post inflationary epoch, since our spectrum is evaluated just at re-entry.

Let us now follow the same procedure for the other branches of the spectrum.
Consider the modes crossing the horizon during the second inflationary phase, with $k_s/c_s<k<k_1$, and characterized by the conditions
\beq
x_s>1,~~~~~~~ c_s x_s>1, ~~~~~~ x_1<1,~~~~~~ c_sx_1<1.
\label{227}
\eeq
For the coefficients $c_\pm^2$ we get exactly the same result as before, given in Eq. (\ref{221}). For the coefficients $c_\pm^3$ we have to use now the small argument limit for all Hankel functions appearing in Eqs. (\ref{217}) and (\ref{218}), and we get:
\beq
\left|c_\pm^3\right| \sim  (c_s x_1)^{-|\nu_2|-1/2}\,x_1 ^{-|\nu_3|+1/2}
\label{228}
\eeq
(modulo numerical factors of order one). This is different from the previous result (\ref{223}), but the spectrum (\ref{29}) is the same. Indeed, when evaluating the perturbation modes at re-entry, in the regime $x_1<1$, we get from Eq. (\ref{28}) 
\beq
\left|\car_k^3\right|_{|k\eta|\sim 1} \sim {1\over \Mp} {\left|c_\pm^3\right|\over \sqrt k}~x_1 ^{-\nu_3+1/2}~.
\label{229}
\eeq
We should now recall that the third phase of decelerated, post-inflationary evolution is characterized by a metric pump field with a standard kinematic power $\a_3>1/2$ (for instance, $\a_3=1$ for radiation and $\a_3=2$ for matter-dominated evolution). Hence, $\nu_3=1/2-\a_3<0$, and $|\nu_3|=-\nu_3$. So, by combining Eqs. (\ref{228}) and (\ref{229}) we find that the $\nu_3$ dependence cancels out, as expected, and we are lead to the same power spectrum as before:
\beq
P(k) \sim  \left(H_1\over \Mp\right)^2 
\left(k\over k_1\right)^{3-2|\nu_2|} c_s^{-1-2|\nu_2|}~,
~~~~~~~~~~~~~~ k_s/c_s<k<k_1.
\label{230}
\eeq

Consider now the intermediate frequency modes of the branch $k_s<k<k_s/c_s$ (typical of a model with $c_s<1$), and characterized by the conditions
\beq
x_s>1,~~~~~~~ c_s x_s<1, ~~~~~~ x_1<1,~~~~~~ c_sx_1<1.
\label{231}
\eeq
An approximate evaluation of the coefficients $c_\pm^2$ now requires the large argument limit (\ref{27}) for the Hankel functions of $x_s$, and the small argument limit (\ref{227}) for the Hankel functions of $c_sx_s$. By taking into account the leading and next-to-leading contributions we can write the result as:
\beq
e^{\pm i\pi\nu_2}\,c_\pm^2= x_s^{1/2} \left[ \a_\pm (c_sx_s)^{|\nu_2|}+\b 
(c_sx_s)^{-|\nu_2|} \right],
\label{232}
\eeq
where $\a_\pm$ and $\b$ are  complex numbers (depending on $\nu_1$ and $\nu_2$), with modulo of order one.

It is important to note that the coefficient $\b$ of the leading term is the same for both $c_+^2$ and $c_-^2$, and  it is essential to include into the computation the subleading contributions (see also \cite{40a}).  
Indeed, by inserting the above result into Eqs. (\ref{217}) and (\ref{218}) for $c_\pm^3$, and using the small-argument expansion (\ref{22}) for the Hankel arguments $x_1$ and $c_sx_1$, one finds that the leading contribution proportional to $(c_sx_s)^{-|\nu_2|}(c_sx_1)^{-|\nu_2|}$ cancels out (because of the factorization of the coefficient $\b$), and we are lead to
\beq
\left|c_\pm^3\right| \sim \left(x_s\over x_1\right)^{|\nu_2|+1/2}\,x_1 ^{-|\nu_3|+1/2}.
\label{233}
\eeq
Using this result into Eq. (\ref{229}) we find as before that the $\nu_3$ contribution drops out, because of the simplification $|\nu_3|+\nu_3=0$, and we obtain the power spectrum
\beq
P(k) \sim  \left(H_1\over \Mp\right)^2 
\left(k_s\over k_1\right)^{3-2|\nu_2|}\left(k\over k_s\right)^4~,
~~~~~~~~~~~~~~ k_s<k<k_s/c_s,
\label{234}
\eeq
which matches continuously with the spectrum (\ref{230}) at $k=k_s/c_s$. Again, as expected, there is no dependence on $\nu_3$, i.e. on the kinematics of the post-inflationary epoch.

Finally, let us consider the lowest frequency modes with $k<k_s$, crossing the horizon during the initial phase of inflation, and characterized by the conditions
\beq
x_s<1,~~~~~~~ c_s x_s<1, ~~~~~~ x_1<1,~~~~~~ c_sx_1<1.
\label{235}
\eeq
For such modes we can insert the small argument limit of all Hankel functions in the computation of both $c_\pm^2$ and $c_\pm^3$. For $c_\pm^2$ we have to take into account also the subleading terms, as before, and the result can be written as
\beq
e^{\pm i\pi\nu_2}\,c_\pm^2= x_s^{-|\nu_1|} \left[ \a_\pm (c_sx_s)^{|\nu_2|}+\b 
(c_sx_s)^{-|\nu_2|} \right],
\label{236}
\eeq
where $|\a_\pm| \sim 1$ and $|\b| \sim 1$. In the computation of $c_\pm^3$ there is again a cancellation of the leading terms, and we obtain
\beq
\left|c_\pm^3\right| \sim {x_s^{-|\nu_1|}\over \sqrt{x_1}}
\left(x_s\over x_1\right)^{|\nu_2|}\,x_1 ^{-|\nu_3|+1/2}.
\label{237}
\eeq
By combing this result with Eq. (\ref{229}), and using again the condition $|\nu_3|+\nu_3=0$, we are lead to the spectrum
\beq
P(k) \sim  \left(H_1\over \Mp\right)^2 
\left(k_s\over k_1\right)^{3-2|\nu_2|}\left(k\over k_s\right)^{3-2 |\nu_1|}~,
~~~~~~~~~~~~~~ k<k_s,
\label{238}
\eeq
which matches continuously with Eq. (\ref{234}) for $k=k_s$ (see also \cite{40a} for a detailed discussion of how the spectral amplitude is affected by the background changes outside the horizon). 

In conclusion, we have found that the complete spectral distribution of all perturbation modes, amplified in the given model of background and evaluated at horizon re-entry,  can be approximately described by our set of equations (\ref{226}), 
(\ref{230}), (\ref{234}) and (\ref{238}). In the following section such results will be applied to a class of models of pre-big bang inflation.

%%%%%%%%%%%%%%%%%%%%%%%%%%%%%%%%%%%%%%%%%%%%%%%
\section{A class of pre-big bang models and the associated perturbation spectra}
\label{Sec3}
\setcounter{equation}{0}
%%%%%%%%%%%%%%%%%%%%%%%%%%%%%%%%%%%%%%%%%%%%%%%

The class of string cosmology models we will consider here is just the same  introduced in recent papers to discuss the production of an observable cosmic background of relic gravitational waves \cite{25}, and the production of primordial seeds for the cosmic magnetic fields \cite{26}. 

The accelerated evolution of such backgrounds starts from an asymptotically flat configuration approaching the string perturbative vacuum \cite{38} at $\eta=-\infty$, and is initially described by a low-energy (but possibly higher-dimensional) phase in which the dilaton $\phi$, the four-dimensional curvature scale $H=a'/a^2$ and the string coupling $g$ are monotonically growing. The initial Kalb-Ramond axion background is trivial, $\sg=0$, but its quantum fluctuations $\da \sg$ are non-vanishing. The dynamical evolution of all these background fields is rigidly prescribed by the low-energy (low-curvature, small coupling) limit of the superstring effective action \cite{16,17,21}.

Let us consider, in particular, a simple example of background geometry with three isotropically expanding dimensions with scale factor $a(\eta)$, and six (internal) space-like shrinking dimensions, not necessarily isotropic, with scale factors $b_i(\eta)$, $i=1, \dots 6$. Working in the so-called string frame (S-frame) parameterization of the string effective action, the pump field $z_h$ controlling the amplification of the four-dimensional (scalar and tensor) metric fluctuations (see e.g. \cite{27,28}), and the pump field $z_\sg$ for the axion fluctuations (see e.g. \cite{29,30}), are determined by the time evolution of the S-frame metric  and of the string coupling $g(\eta)$ as follows:
\beq
z_h(\eta) \sim a g^{-1} ~, ~~~~~~~~~~~~
z_\sg(\eta) \sim a g~, ~~~~~~~~~~~~
g(\eta) \sim \left(\prod_{i=1}^6 b_i  \right)^{-1/2} e^{\phi/2}
\label{31}
\eeq
(we have included into the effective four-dimensional string coupling $g$ the possible dynamical contribution of the internal volume, if not frozen). 

Using the explicit solutions of the string cosmology equations, and parameterizing the pump fields as in Eq. (\ref{23}), we then find that during the first, low-energy inflationary phase, the behavior of $z_h(\eta)$ depends on the power $\a_1$ rigidly fixed such that \cite{27,28}:
\beq
\a_1={1/2}\,, ~~~~~~~~ \nu_1= {1/2} - \a_1=0\,, ~~~~~~~~
3-2|\nu_1|=3\,.
\label{32}
\eeq
The power-law behavior of the axion pump field $z_\sg(\eta)$, on the contrary, depends on the parameters which control the isotropy of the background geometry, and may give a perfectly scale-invariant spectrum in the limit of exact isotropy and duality symmetry between internal and external dimensions \cite{16,21,39}. Following \cite{25,26,39} we shall thus fix the power-law parameter $\a_1$ of $z_\sg$ in such a way as to allow a small ``red" tilt of the large-scale  perturbations sourced by the axion, i.e. such that 
\beq
3-2|\nu_1|= 3-|1- 2\a_1| \equiv n_s-1.
\label{33}
\eeq
We have used the standard notation for the scalar spectral index $n_s$ (see e.g. \cite{40}) where, in agreement with recent observational data \cite{40}, we take $n_s \simeq 0.965$.

This first inflationary regime ends at the epoch $\eta=-\eta_s$ when the S-frame background curvature reaches the string scale. At that point, even if all background fields are well inside the weak coupling regime ($g \ll 1$), a consistent dynamics requires the inclusion of the high-curvature $\ap$ corrections into the string effective action. As a consequence we obtain modified cosmological equations, and the background tends to stabilize on a second phase of inflation, described by a solution which possibly represents a late-time attractor of the preceding low-energy phase \cite{41} (see also \cite{42}, and see \cite{43,44,45} for the possibility of non-perturbative solutions to all orders in $\ap$). 

During this second high-energy inflationary phase, ranging from $\eta=-\eta_s$ to $\eta=-\eta_1$, the definition of the pump fields (\ref{31}) is still valid, but now the S-frame curvature $H=a'/a^2$ tends to stay constant, so that $a(\eta) \sim (-\eta)^{-1}$, while the effective string coupling keeps growing \cite{41}. Such a growth can be parameterized by a simple power-law behavior, in conformal time, as $g(\eta) \sim (-\eta)^{-\b}$, where $\b$ is an unknown (positive) parameter. Hence, for the metric pump field $z_h$ of eq. (\ref{31}), the power $\a_2$ now satisfies
\beq
\a_2=-1+\b, ~~~~~~~~ \nu_2= {3/2} - \b, ~~~~~~~~
3-2|\nu_2|=3- |3-2\b|.
\label{34}
\eeq 
For the axion pump field $z_\sg$ we have instead, from Eq. (\ref{31}), a corresponding power $\a_2$ such that
\beq
\a_2=-1-\b, ~~~~~~~~ \nu_2= {3/2} + \b, ~~~~~~~~
3-2|\nu_2|=3- |3+2\b|.
\label{35}
\eeq 
In the rest of this paper we will always keep the absolute value in the above (and similar) equations, to allow for possible negative values of the argument $3 \pm 2\b$  (however, in this paper, we will limit to the case $\b>0$, i.e. growing dilaton). Notice that a non-trivial evolution of both the dilaton and the internal dimensions, during the initial low-energy phase as well as the final high-curvature string phase, is crucial to fix the final spectral behavior of both metric and axion field perturbations.

This second regime of inflationary pre-big bang evolution eventually ends at the epoch $\eta=-\eta_1$, when the background enters the strong coupling regime (with $g_1 \equiv g(-\eta_1) \sim 1$), and we may expect a rapid ``bouncing" transition -- possibly due to the contributions of a non-local dilaton potential \cite{46,47} -- to the  post-big bang regime of standard, decelerated evolution. 
It should be noted that the dynamical details of the bounce can in principle contribute to modify the sound velocity $c_s$ and thus affect the (very high) frequency modes crossing the horizon during the bouncing
phase. For the applications of this paper we will
neglect such a possible distortion of the spectrum near the end-point-frequency, assuming that the bounce is ``almost instantaneously" localized at the transition epoch $\eta= -\eta_1$.

We shall assume, also, that the transition occurs at a curvature scale $H_1$ smaller than Planckian ($H_1<\Mp = 1/\sqrt{8\pi G}$), and that in the post-big bang epoch the dilaton and the extra spatial dimensions are frozen. 
This occurs because, in our class of models, the stabilization of the dilaton and of the internal dimensions is assumed to be closely related to the mechanism responsible for the bouncing transition, and thus fully localized at the end of the string phase, as stressed before.
Most importantly, we shall assume that the post-big bang axion background has the right properties to implement the curvaton mechanism, so as to ``translate" the  primordial spectrum of isocurvature axion perturbations into a final spectrum of adiabatic (and Gaussian) scalar curvature perturbations \cite{23,24}. 

What we need, to this purpose, is that the axion background emerges from the bounce with a mass $m_\sg$ and a non-trivial value $\sg_i \not=0$ displaced from the minimum of the (non-perturbative, periodic) axion potential, see e.g.  \cite{23,24}. Excluding the exotic possibility of trans-Planckian values $\sg_i>\Mp$, we shall adopt here the simple choices $\sg_i \sim \Mp$ and $m_\sg \sim H_1$, which lead us to the class of models already considered in \cite{25,26}. 

For such models the axion background starts oscillating just immediately after inflation, at the epoch $\eta=-\eta_1$; it  then behaves like a dust fluid, which dominates the cosmological evolution until it eventually decays (into electromagnetic radiation, with gravitational coupling strength) at the decay epoch $\eta_d$, corresponding to  the decay scale $H_d \sim Gm_\sg^3 \sim H_1^3/ \Mp^2$. After that scale (i.e. for $\eta >\eta_d$) the Universe enters a final radiation-dominated regime.

The fact that the axion decay occurs while the axion is dominant, in particular, avoids the possible introduction of ``non-Gaussian" properties in the perturbation spectrum \cite{48}. In any case, the decay automatically produces a super-horizon spectrum of adiabatic curvature perturbations described by the gauge-invariant (scalar) variable $\car$, whose spectral modes, $\car_k$, are in general related to the axion modes, $\da \sg_k$, by the simple transfer function $f(\sg_i)$ such that: 
\beq
\left|\car_k\right| \simeq f(\sg_i) \left| \da \sg_k\right|, ~~~~~~~~~~~~~~~
f(\sg_i) \simeq 0.13{\sg_i\over \Mp} + 0.25 {\Mp\over \sg_i} -0.01.
\label{36}
\eeq
The analytical and numerical expression for $f(\sg_i)$ has been derived in 
{\cite{23,24}. For 
our  case  $\sg_i=\Mp$ we obtain, in particular, $f^2(\sg_i) \simeq 0.137$. 

We are now in the position of introducing the explicit spectral distributions of scalar, tensor and axion-induced curvature perturbations amplified by our model of pre-big bang inflation, and emerging in a regime of axion-dominated or radiation dominated post-big bang evolution. As anticipated in Sect. \ref{Sec1},  we will take into account a possible modification of the sound speed parameter ($c_s<1$) during the high-curvature string phase, but only for the evolution of the axionic perturbations.

By exploiting the results of Sect. \ref{Sec2}, let us first express the primordial spectrum of scalar curvature perturbations due to the axion, and evaluated at horizon re-entry. By using 
Eqs. (\ref{226}), (\ref{230}), (\ref{234}), (\ref{238}), (\ref{33}), (\ref{35}) and (\ref{36}) we obtain:
\bea
P_\sg(\om) &\simeq&  {f^2\over 2\pi^2 g_1^4}\left(H_1\over \Mp\right)^2 
\left(\om\over \om_1\right)^{3-|3+2\b|} c_s^{-1-|3+2\b|}~,
~~~~~~~~~~ {\om_s\over c_s}<\om<{\om_1\over c_s},
\nonumber \\ &\simeq&
{f^2\over 2\pi^2 g_1^4} \left(H_1\over \Mp\right)^2 
\left(\om_s\over \om_1\right)^{3-|3+2\b|}\left(\om\over \om_s\right)^4~,
~~~~~~~~~~~~~~ \om_s<\om<{\om_s\over c_s},
\nonumber \\ &\simeq&
{f^2\over 2\pi^2 g_1^4} \left(H_1\over \Mp\right)^2 
\left(\om_s\over \om_1\right)^{3-|3+2\b|}\left(\om\over \om_s\right)^{n_s-1}~,
~~~~~~~~~~\om<\om_s,
\label{37}
\eea
We have followed the canonical definition (\ref{29}) using however, for later convenience, the proper frequencies $\om(t)= k/a(t)$ instead of the comoving frequencies $k$. Possible numerical factors of order one have been absorbed into the curvature-scale parameter $H_1$. The factor $f^2$ is due to Eq. (\ref{36}). Finally, let us explain in detail the presence of the overall factor $g_1^{-4}$ in the spectral amplitude. 

Let us recall, to this purpose, that we have worked with the pump fields (\ref{31}) defined in the S-frame, and that, in such a context, the metric pump field $z_h(-\eta_1) \equiv a_1 g_1^{-1}$ exactly coincides with the Einstein frame (E-frame) scale factor
$\ti a(\eta) \equiv  a g^{-1}$ (see e.g. \cite{21}), evaluated at $\eta= -\eta_1$. The spectral normalization adopted in Eq. (\ref{226}) for metric perturbations, and based on the unit values of the scale factor at the transition epoch, thus directly refers to the curvature scale $H_1$ of the E-frame geometry, namely to the quantity $k_1/ \ti a_1 \sim \ti H_1$. On the other hand, the axion pump field of Eq. (\ref{31}) gives $z_\sg(-\eta_1) = a_1g_1= \ti a_1 g_1^2$, so that the corresponding normalization at the end of inflation, for the axion spectrum, leads to the parameter $k_1/(\ti a_1 g_1^2) \equiv \ti H_1/g_1^2$. Hence, if we want to express the spectral amplitude of axion and metric perturbations in terms of the same (E-frame) curvature scale $\ti H_1$ (omitting the tilde, for simplicity), we have to insert into the axion spectrum also the appropriate dependence on the final value of the string coupling $g_1$ (even if we expect, in a realistic scenario, a value not very different from the natural choice $g_1 \sim 1$).  

To the axion-induced curvature perturbations we have now to add the primordial spectrum $P_\psi$ of the scalar metric perturbations described by the gauge-invariant variable $\psi_k$, and directly amplified by the two phases of inflation. Using the pump-field powers (\ref{32}), (\ref{34}), and following the same procedure as before (but setting everywhere $c_s=1$), we find that the corresponding power-spectrum, evaluated at re-entry, is given by
\bea
P_\psi(\om) &\simeq&  {1\over 2\pi^2}\left(H_1\over \Mp\right)^2 
\left(\om\over \om_1\right)^{3-|3-2\b|}~,
~~~~~~~~~~~~~~~~~~~~~~ {\om_s}<\om<{\om_1},
\nonumber \\ &\simeq&
{1\over 2\pi^2 } \left(H_1\over \Mp\right)^2 
\left(\om_s\over \om_1\right)^{3-|3-2\b|}\left(\om\over \om_s\right)^{3}~,
~~~~~~~~~~~~~\om<\om_s.
\label{38}
\eea
For the lowest frequency modes, crossing the horizon during the first, low-energy phase of inflation, we thus recover the well known (and very steep) cubic tilt of the spectrum \cite{27,28} (modulo logarithmic corrections, not included here for simplicity).

Finally, our model of inflation also amplifies tensor metric perturbations and produces a relic background of cosmic gravitational radiation, which can be conveniently described by its spectral energy density $\Om_g$ expressed in critical units and evaluated at the present time $t_0$, i.e. $\Om_g(\om. t_0)= \rho_g(\om, t_0)/\r_c(t_0)$. 

The pump field is the same as that of scalar metric perturbations, but there is a difference due to the fact that, for tensor perturbations, we are not evaluating the primordial spectrum at re-entry, but we are  interested in the today value of their spectral distribution, thus referring to perturbation modes which are ``well inside" the horizon. This means that the spectral distribution may be in principle affected by the sub-horizon evolution of the modes, after their re-entry. In  our class of models there is indeed a different spectral behavior for modes re-entering the horizon during the radiation-dominated or the axion-dominated epoch. In the first case the slope of the primordial spectrum is unchanged, in the second case the slope is tilted towards the red by an additional $\om^{-2}$ factor (a well know effect of the phases of dust-dominated evolution \cite{16,21}, like the one dominated by an oscillating axion).

Taking into account this effect for the band of high-frequency modes with $\om>\om_d$, re-entering before the axion decay (i.e. at a  curvature scale $H>H_d$), the produced spectral energy density of relic gravitons (again, putting everywhere $c_s=1$) can then be written as follows (see also \cite{25,26}): 
\bea
\Om_{g}(\om, t_0) &\simeq& 
\Om_r (t_0) \left(H_1\over \Mp\right)^{10/3} 
\left(\om\over \om_1\right)^{1-|3-2\b|},
~~~~~~~~~~~~~~~~~~~~~~~~~~~~~~~~ \om_d<\om<\om_1,
\nonumber \\ &\simeq&
\Om_r (t_0) \left(H_1\over \Mp\right)^{10/3}
\left(\om_d\over \om_1\right)^{1-|3-2\b|}
\left(\om\over \om_d\right)^{3-|3-2\b|},
~~~~~~~~~~~~~~ \om_s<\om<\om_d,
\nonumber \\ &\simeq&
\Om_r (t_0) \left(H_1\over \Mp\right)^{10/3}
\left(\om_d\over \om_1\right)^{1-|3-2\b|}
 \left(\om_s\over \om_d\right)^{3-|3-2\b|} \left(\om\over \om_s\right)^3,
~~~~~ \om<\om_s.
\label{39}
\eea
Here $\Om_r(t_0) \sim 10^{-4}$ is the present fraction of critical energy density in the form of cosmic radiation, and we have assumed that $\om_d>\om_s$, namely that the initial post-big bang phase of axion-dominated evolution is short enough to affect the spectrum only for the high-frequency modes leaving the horizon during the high-curvature string phase.  Finally, we have not included here the lowest frequency band of the modes re-entering the horizon very late, after equality ($\om <\om_{\rm eq}$) because, given the very steep blue tilt of the other branches, its amplitude turns out to be too small to have significant effects for the discussion of this paper.

It should be noted, finally, that the above spectrum has already been presented and discussed in previous papers \cite{25,26}, with the only difference that here (for the sake of accuracy) we have explicitly taken into account the possible contribution of the axion energy density $\r_\sg$ to the total critical density $\r_c$ during the initial regime of axion dominance. This rescales (actually, slowly depress) the overall spectral amplitude by the constant factor $(H_1/\Mp)^{4/3}$, and leads to an overall dependence on $H_1/\Mp$ slightly different from that reported for the other primordial spectra. We have checked, however, that this effect has no crucial impact at all on our following phenomenological discussion, and on the qualitative results of  this paper.

%%%%%%%%%%%%%%%%%%%%%%%%%%%%%%%%%%%%%%%%%%%%%%%
\section{Phenomenological and self-consistency constraints}
\label{Sec4}
\setcounter{equation}{0}
%%%%%%%%%%%%%%%%%%%%%%%%%%%%%%%%%%%%%%%%%%%%%%%

We are now ready to impose on the spectral distributions (\ref{37}), (\ref{38}) and (\ref{39}) the appropriate constraints arising from phenomenological as well as self-consistency conditions, in order to explore the allowed region of parameter space.

Let us notice, to this purpose, that the spectra produced in our class of models depend (after fixing the numerical values of $n_s$ and $f^2(\sg_i)$) on four arbitrary parameters: $\eta_s$, $\eta_1$, $\b$ and $g_1$. However, as already anticipated, our discussion will be limited to models with $g_1$ fixed to the (natural) value $g_1=1$. We are thus left with three free parameters, and we can note that $\eta_s$ and $\eta_1$, determining the duration and the localization in time of the high-energy string phase, can be conveniently replaced by two (more physical) quantities: the final (E-frame) curvature scale $H_1$, and the redshift parameter
\beq
z_s= \eta_s/\eta_1= a_1/a_s = \om_1/\om_s,
\label{41}
\eeq
associated with the expansion of the three-dimensional space during the string phase (the last equality, in the above equation, is due to the constancy of the S-frame Hubble parameter during the string phase). Consequently, the third parameter $\b$ can be expressed in terms of the overall growth of the effective coupling $g$ during the string phase as:
\beq
g_s/g_1 = (\eta_s/\eta_1)^{-\b} = z_s^{-\b}.
\label{42}
\eeq

As will be shown below, the parameter $H_1$ turns out to be fixed in term of $\b$ and $z_s$ by the constraint imposed by the CMB normalization of the scalar spectrum, i.e. $H_1= H_1(\b, z_s)$: using that relation, it will be possible to eliminate $H_1$ everywhere. In this way we end up with a two-dimensional parameter space, which can be conveniently parametrized (mainly for graphical reasons) by the two logarithmic variables $\{x, y\}$ defined by:
\beq
x = \log z_s, ~~~~~~~~~~~~~~~~~~~~~~ 
y= \log (g_s/g_1) = -\b x
\label{43}
\eeq
(see also \cite{25,26}). Let us stress, finally, that in the list of constraints that we are going to introduce same of them are redundant, in the sense that they may be an implicit consequence of other, stronger constraints. However, it will be useful to present here the full list of conditions in view of possible (future) discussions of more general background models. 

Let us start by noting that the range of the parameter $\b$ is in principle constrained by the conditions $0 \leq \b \leq 3$, where the lower limit is due to our assumption of growing string coupling, while the upper limit is to be imposed to avoid background instabilities \cite{49}. Hence, according to our definitions (\ref{43}), 
\beq
-3 x < y<0.
\label{44}
\eeq
Second, in order to express $H_1$ in terms of the other two parameters, we shall require that the lowest frequency branch ($\om <\om_s$) of the axionic spectrum (\ref{37}), when evaluated at the proper frequency $\om_*$ corresponding to the so-called {\it pivot} scale $k_*= 0.05$ Mpc$^{-1}$, exactly satisfies  the normalization imposed by the CMB data \cite{40}, namely $P_\sg(\om_*) \simeq 3 \times 10^{-10}$. This implies, using the definition (\ref{41}),
\beq
{f^2\over 2\pi^2 g_1^4}\left(H_1\over \Mp\right)^2  z_s^{|3+2\b|-3}
\left(\om_*\over \om_s\right)^{n_s-1} \simeq 3  \times10^{-10}.
\label{45}
\eeq
On the other hand, an explicit computation of the ratio $(\om_*/ \om_s)$ using the standard background evolution (from the initial scale $H_1$ down to $H_*$), gives \cite{25}
\beq
\left(\om_*\over \om_s\right) = z_s \left(\om_*\over \om_1\right)=
z_s{ H_* a_*\over H_1 a_1} \simeq 10^{-27} z_s \left(H_1\over \Mp\right)^{-5/6},
\label{46}
\eeq
where we have used \cite{40} $H_* \simeq 2 \times 10^{-54} \Mp$. According to the definitions (\ref{43}) we can then write the constraint (\ref{45}) in a convenient logarithmic form as
\beq
\log \left(H_1\over \Mp\right) \simeq {6\over 17- 5 n_s} \left[ (4-n_s)x - |3x -2y| + 27 n_s - 37 + \log\left( 6 \pi^2 g_1^4\over f^2\right) \right],
\label{47}
\eeq
which will allow us to eliminate everywhere $H_1$ in terms of $x$ and $y$. 

We can then easily impose the assumed condition that the bouncing transition occurs at a given (E-frame) curvature scale smaller than Planckian, i.e.
\beq
\log \left(H_1\over \Mp\right)<0.
\label{48}
\eeq
To this, we have to add the condition of that the produced scalar perturbations have a negligible backreaction on the assumed model of background 
evolution\footnote{Including the  backreaction of perturbations could change in a non-trivial way the background dynamics. We plan to study such backreaction effects  in a forthcoming paper using the covariant approach introduced in \cite{Gasperini:2009wp,Gasperini:2009mu,Marozzi:2010qz}, and already applied in 
perturbative (see for example \cite{Finelli:2011cw,Marozzi:2012tp}) as well as non-perturbative  (see \cite{Brandenberger:2018fdd}) calculations.}, i.e. $P_\psi<1$, $P_\sg<1$. Such conditions are to be imposed at the peak values of the power spectra (\ref{37}) and (\ref{38}), located at $\om=\om_1$ for $P_\psi$ and at $\om=\om_s/c_s$ for $P_\sg$. This leads, respectively, to
\bea
&&
\log \left(H_1\over \Mp\right)<{1\over 2} \log (2 \pi^2),
\nonumber \\ &&
\log \left(H_1\over \Mp\right) <2 \log c_s +{1\over 2} \log \left (2 \pi^2 g_1^4\over f^2\right)+ {1\over 2} \left(3x- |3x-2y| \right),
\label{49}
\eea
and this second constraint is in general more stringent, for $c_s<1$, than the one of Eq. (\ref{48}). 

We have also to add a series of conditions following from the assumed  hierarchy of scales, typical of our class of models. 

First of all we have to impose that all frequency scales up to the highest one ($\om_{LS}\sim 3$ Mpc$^{-1}$) constrained by  Large Scale Structures  (LSS) observations still belong to the lowest frequency branch of the axionic spectrum (\ref{37}), namely that $\om_{LS}<\om_s$. This condition is required to match present observations with a flat enough, slightly ``red" primordial scalar spectrum. By using the numerical relation $\om_{LS}  \simeq 60 \,\om_*$ and the result of Eq. (\ref{46}) we then obtain the condition
\beq
\left(\om_{LS}\over \om_s\right)  \simeq 6 \times10^{-26} z_s \left(H_1\over \Mp\right)^{-5/6}<1,
\label{410}
\eeq
namely 
\beq
x- {5\over 6}\log \left(H_1\over \Mp\right)<26-\log 6.
\label{411}
\eeq
In addition, we have to impose that, up to the scale $\om_{LS}$, the low-energy (flat) axionic spectrum (\ref{37}) dominates over the (very blue) low-frequency band of the scalar spectrum (\ref{38}), namely that $P_\psi(\om_{LS})<P_\sg(\om_{LS})$. This implies, using Eqs. (\ref{43}), (\ref{47}) and (\ref{410}),
\beq
|3x+2y|-|3x-2y|+ (4-n_s) \left[x-26+\log 6 -{5\over 6} \log \left(H_1\over \Mp\right)\right] < \log \left(f^2\over g_1^4\right).
\label{412}
\eeq

Finally, concerning the important axion-decay scale, and in order to avoid disturbing the standard nucleosynthesis scenario with a dust-dominated phase, we have to impose that the decay  of the oscillating axion occurs at the scale $H_d \sim H_1^3/\Mp^2$ preceding the nucleosynthesis scale $H_N \sim (1\, {\rm MeV})^2/\Mp$, i.e.  $H_d > H_N$, which implies
\beq
 \log \left(H_1\over \Mp\right)\gaq -14.
 \label{413}
 \eeq
 We should also recall that, in the previous section, we have assumed that the axion-dominated phase is short enough to affect only the highest frequency branch of the tensor perturbation spectrum,  namely that $\om_d>\om_s$. Let us be here slightly more restrictive (also in view of possible future generalizations), by assuming that this is true for all types of perturbation spectra. Hence, considering in particular the axionic spectrum (\ref{37}), we shall assume the stronger condition $\om_d > \om_s/c_s$, namely
\beq
{\om_s\over \om_d} = z_s^{-1} {\om_1\over \om_d}=
z_s^{-1}{ H_1a_1\over H_d a_d} = z_s^{-1}  \left(H_1\over \Mp\right)^{-2/3}<c_s,
\label{414}
\eeq
which implies, in logarithmic form,
\beq
 \log \left(H_1\over \Mp\right)>-{3\over 2} \left(x+\log c_s\right).
 \label{415}
 \eeq

We should consider now the phenomenological constraints imposed by the presence of a relic background of cosmic gravitons, described by the spectral energy density (\ref{39}). 

A first condition come from nucleosynthesis \cite{50}: the present value of the total gravitational energy density $\Om_g(t_0)$, integrated over all modes and rescaled down to the nucleosynthesis epoch, cannot exceed (roughly) the energy density of one massless degrees of freedom in thermal equilibrium (see e.g. \cite{51} for a detailed computation). This bound can be translated into a crude upper limit on the peak intensity of the spectrum, i.e. $\Om_g(\om_{\rm peak}) < 10^{-1} \Om_r$, to be imposed at the peak frequency $\om_{\rm peak}$ which, for our spectrum (\ref{39}), may correspond to either  $\om_1$ or $\om_d$, depending on the value of $\b$. In the first case we are lead to the condition 
\beq
 \log \left(H_1\over \Mp\right)<-{3\over 10} .
 \label{416}
 \eeq
In the second case to a condition which, using the relation $\om_d/\om_1= (H_1/\Mp)^{2/3}$, can be written as:
\beq
{1\over 3} \left(12 -\left|6+4{y\over x}\right| \right)  \log \left(H_1\over \Mp\right)<-1.
 \label{417}
 \eeq
 
 A possible more stringent upper bound on $\Om_g$ has been recently placed by the Advanced LIGO and Virgo first observing run \cite{52}, namely $\Om_g(\om_L) <10^{-7}$, at the frequency scale typical of the sensitivity of presently operating interferometric detectors, $\om_L \sim 10^2 Hz$. To impose such a limit we have to distinguish three cases, depending on the localization of $\om_L$ in the various frequency branches of our tensor spectrum (\ref{39}). 
 
 We recall, to this purpose, that $\om_* \sim 5 \times 10^{-16}$ Hz so that $\om_L \simeq 0.2 \times 10^{18} \om_*$ and, by using our previous results (\ref{46}) and (\ref{414}), we obtain:
 \beq
{\om_L\over \om_d} = 2\times 10^{-10} \left(H_1\over \Mp\right)^{-3/2},
~~~~~~~~~
{\om_L\over \om_s}= z_s{\om_L\over \om_1}=
 2\times 10^{-10} z_s\left(H_1\over \Mp\right)^{-5/6}.
\label{418}
\eeq
By imposing the LIGO constraints in the three possible spectral sectors, and comparing the result with all  previously imposed conditions, we find, as the only new relevant restriction, that $\om_L$ must satisfy the condition $\om_L<\om_1$, namely
\beq
 \log \left(H_1\over \Mp\right)>-12 + {6\over 5}  \log 2.
 \label{419}
 \eeq

A final (in principle important) phenomenological  constraint for the graviton spectrum comes from the observations of millisecond pulsars (see e.g. \cite{53}), which give the bound $\Om_g(\om_p) <10^{-8}$, at the frequency scale $\om_p \sim 10^{-8}$  Hz $\sim 0.2 \times 10^8\om_* \sim 10^{-10} \om_L$. As before, such a constraint has to be  separately imposed on  the three different branches of the graviton spectrum (\ref{39}), depending on the localization of $\om_p$ with respect to $\om_s$, $\om_d$ and $\om_1$. By using the above relations (\ref{418}), appropriately rescaled for $\om_L \ra \om_p$, and imposing the corresponding spectral constraints (that will not be explicitly   reproduced here), we find that there are no further restrictions on our  parameters, stronger than the ones already reported.

Summarizing, and applying all the relevant constraints listed in this Section, we can now display the allowed region in the two-dimensional parameter space of our class of models, in the plane spanned by the variables $x$ and $y$ controlling, respectively, the temporal extension of the string phase and the associated growth of the string coupling (see eq. (\ref{43})). 

We find that there is in principle a wide allowed region, but that such a region is  quite sensitive to the effective value of the  sound-speed parameter when $c_s <<1$. Indeed, for small values of $c_s$ the amplitude of $P_\sg$ tends to be enhanced (see Eq. (\ref{37})), and the constraint $P_\sg<1$ starts to play a crucial role (see Eq. (\ref{49})). In practice, the region allowed by our set of spectral constraints is almost insensitive to $c_s$ for values ranging from $c_s=1$ down to $c_s \simeq 0.004$, but it practically disappears for $c_s\laq 0.003$. This effect is illustrated in in Fig. \ref{f1}, where we have plotted the allowed region for the three values $c_s=1$, $c_s =0.0038$ and $c_s=0.0032$. 

%%%%%%%%%%%%%%%%%%%%%%%%%%%%%%%%%%%%%%%%%%%%%%%%%%%%%%%%%%%%%%%%%%%%%%%%%%%
\begin{figure}[t]
\centering
\includegraphics[width=9cm]{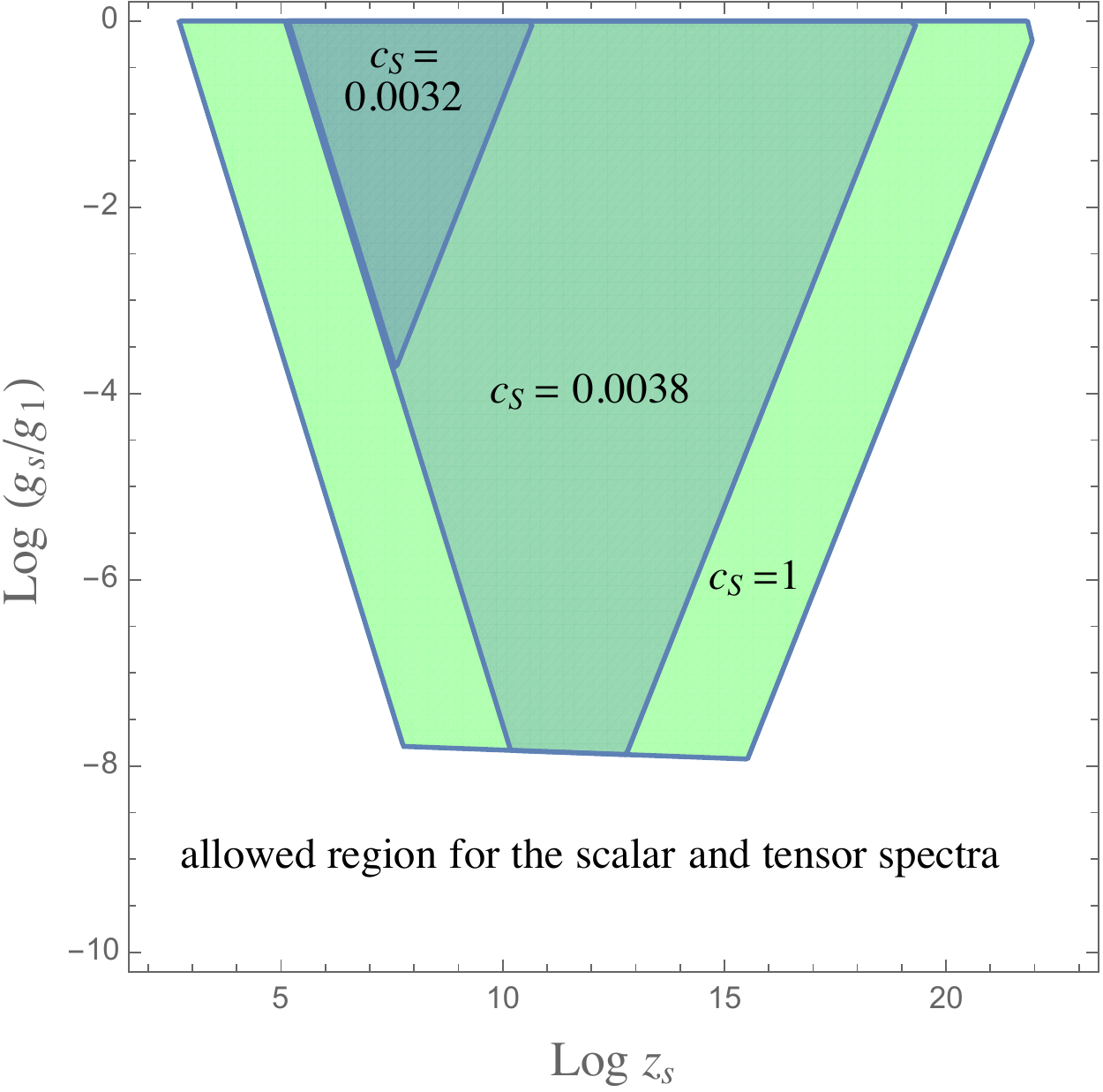}
\caption{The allowed region of parameter space for three different values of the sound-speed parameter $c_s$, controlling the evolution of the axion perturbations during the high-curvature string phase. The region goes rapidly to zero for values of $c_s$ smaller than 0.004. } 
\label{f1}
\end{figure}
%%%%%%%%%%%%%%%%%%%%%%%%%%%%%%%%%%%%%%%%%%%%%%%%%%%%%%%%%%%%%%%%%%%%%%%%%%%

%%%%%%%%%%%%%%%%%%%%%%%%%%%%%%%%%%%%%%%%%%%%%%%
\section{Allowed region for PBH production}
\label{Sec5}
\setcounter{equation}{0}
%%%%%%%%%%%%%%%%%%%%%%%%%%%%%%%%%%%%%%%%%%%%%%%

Let us first recall that the probability of PBH formation via the collapse of big primordial inhomogeneities, re-entering the horizon after inflation, is closely related to the amplitude of the associated scalar perturbations \cite{13}. In particular, a perturbation mode $\om_M$, re-entering at the scale $H_M=\om_M$, can source an interesting abundance of PBHs of mass $M$ of the order of the horizon mass, i.e. 
\beq
M \sim \Mp^2/H_M,
\label{51}
\eeq
provided the spectral amplitude $P(\om_M)$, at horizon crossing, satisfies the condition $P(\om_M)\geq 10^{-2}$ (see e.g. \cite{54}, but see also \cite{germani2} for a possibly more accurate evaluation of the required spectral amplitude). For the purpose of this paper, devoted to a (preliminary) qualitative investigation of PBH production in a string cosmology context, we will thus assume the above condition as our reference criterium to discuss the possibility that PBHs are formed in the physically interesting range of masses $M \sim 10^{18}-10^{22}$ g. 

In order to impose the condition $P(\om_M)\geq 10^{-2}$ on our scalar spectra (\ref{37}) and (\ref{38}), and to explicitly formulate  such condition in terms of the mass parameter $M$ of Eq. (\ref{51}), we must distinguish four cases, depending on the exit and re-enter scale of the mode $\om_M$. The exit scale is needed to localize $\om_M$ in the appropriate branch of the spectrum, while the re-enter scale is needed to appropriately relate $\om_M$ to $M$ according to the given model of post-inflationary evolution. 

For our model we have, in particular, the following possibilities: $i)$ exit in the string phase, re-enter in the axion-dominated phase,
\beq
\om_d < \om_M <\om_1/c_s~;
\label{52}
\eeq
$ii)$ exit in the string phase, re-enter in the radiation phase,
\beq
\om_s/c_s < \om_M <\om_d~;
\label{53}
\eeq
$iii)$ exit just at the beginning of the string phase due to the sudded change of $c_s$, 
re-enter in the radiation phase,
\beq
\om_s < \om_M <\om_s/c_s~;
\label{54}
\eeq
$iv)$ exit in the dilaton phase, re-enter in the radiation phase,
\beq
\om_{LS} < \om_M <\om_s
\label{55}
\eeq
(note that with the condition $\om_M>\om_{LS}$ we are limiting the discussion to frequency scales higher than the typical LSS scale, to avoid entering the range of too large PBH masses, not relevant as possible components of dark matter). 
In each of these four frequency bands we have to take into account the two alternative possibilities in which metric perturbations are dominant, $P_\psi>P_\sg$, or axionic perturbations are dominant, $P_\sg>P_\psi$, and consequently impose the appropriate condition, either $P_\psi(\om_M)>10^{-2}$ or  
$P_\sg(\om_M)>10^{-2}$. Finally, in all cases we have to add the conditions 
$P_\psi<1$, $P_\sg<1$ (already explicitly reported in Eq. (\ref{49})), to exclude background instabilities due to a too large amplification of perturbations. 

Let us start with the highest frequency branch (\ref{52}), for which we need to evaluate $\om_M/\om_1$ and $\om_M/\om_d$. By using eq. (\ref{51}) and our previous result (\ref{414}) we obtain:
\bea
&&
{\om_M\over \om_1} ={ H_M a_M\over H_1 a_1} = \left(H_1\over \Mp\right)^{-1/3}
\left(M\over \Mp\right)^{-1/3}, 
\nonumber \\&&
{\om_M\over \om_d} ={\om_M\over \om_1} {\om_1\over \om_d} 
 = \left(H_1\over \Mp\right)^{-1}
\left(M\over \Mp\right)^{-1/3}. 
\label{56}
\eea
If $P_\psi$ is dominant, in this branch defined by 
\beq
-\log  \left(H_1\over \Mp\right)<\log \left(M\over \Mp\right),
~~~~~~~~~
-\log  \left(H_1\over \Mp\right)>{1\over 3} \log \left(M\over \Mp\right),
\label{57}
\eeq
we have to impose the condition $P_\psi(\om_M)>10^{-2}$, which takes the explicit form
\beq
{1\over 3} \left(3+\left|3+2{y\over x}\right| \right)  \log \left(H_1\over \Mp\right)
-{1\over 3}  \left(3-\left|3+2{y\over x}\right| \right)  \log \left(M\over \Mp\right)
>-2 +\log (2\pi^2).
\label{58}
\eeq
If, on the contrary, the spectrum $P_\sg$ is dominant, then the first condition of Eq. (\ref{57}) has to be replaced by
\beq
-\log  \left(H_1\over \Mp\right)<\log \left(M\over \Mp\right) - 3 \log c_s,
\label{59}
\eeq
and the spectral condition becomes $P_\sg(\om_M)>10^{-2}$, namely:
\bea
&& \!\!\!\!\!\!\!\!\!\!\!\!\!\!\!\!\!\!\!\!\!\!\!\!\!\!\!\!\!\!
{1\over 3} \left(3+\left|3-2{y\over x}\right| \right)  \log \left(H_1\over \Mp\right)
-{1\over 3}  \left(3-\left|3-2{y\over x}\right| \right)  \log \left(M\over \Mp\right)>
\nonumber \\ && ~~~~~~~~~~~~~~~~~~~~~~~
-2 +\log \left(2\pi^2 g_1^4\over f^2 \right) + \left(1+\left|3-2{y\over x}\right| \right)\log c_s.
\label{510}
\eea

The same procedure is to be followed for the other three spectral branches. For the branch of Eq. (\ref{53}) we have 
\bea
&&
{\om_M\over \om_d} ={ H_M a_M\over H_d a_d} = \left(H_1\over \Mp\right)^{-3/2}
\left(M\over \Mp\right)^{-1/2}, 
\nonumber \\&&
{\om_M\over \om_s} =z_s{ H_M a_M\over H_1 a_1}=  z_s  
 \left(H_1\over \Mp\right)^{-5/6}
\left(M\over \Mp\right)^{-1/2}. 
\label{511}
\eea
It is important to note that $\om_M/\om_d$ and $\om_M/\om_1$ in the above equation are both different from the analogous expressions reported in Eq. (\ref{56}),  because now the mode $\om_M<\om_d$ re-enters in the radiation era, where $a_M(\eta) \sim H^{-1/2}$ (differently from the previous case where $\om_M>\om_d$ and $a_M(\eta) \sim H^{-2/3}$). 
If $P_\psi$ is dominant we have, for this branch, 
\beq
-{3\over 2}\log  \left(H_1\over \Mp\right)< {1\over 2}\log \left(M\over \Mp\right),
~~~~~~~~~
x- {5\over 6}\log  \left(H_1\over \Mp\right)>{1\over 2} \log \left(M\over \Mp\right),
\label{512}
\eeq
and the constraint $P_\psi(\om_M)>10^{-2}$gives
\beq
{1\over 6} \left(-3+5\left|3+2{y\over x}\right| \right)  \log \left(H_1\over \Mp\right)
-{1\over 2}  \left(3-\left|3+2{y\over x}\right| \right)  \log \left(M\over \Mp\right)
>-2 +\log (2\pi^2).
\label{513}
\eeq
If the spectrum $P_\sg$ is dominant, then the second condition of Eq. (\ref{512}) has to be replaced by
\beq
x -{5\over 6}\log  \left(H_1\over \Mp\right)>{1\over 2}\log \left(M\over \Mp\right) -  \log c_s,
\label{514}
\eeq
and the spectral condition becomes $P_\sg(\om_M)>10^{-2}$, namely:
\bea
&& \!\!\!\!\!\!\!\!\!\!\!\!\!\!\!\!\!\!\!\!\!\!\!\!\!\!\!\!\!\!
{1\over 6} \left(-3+5 \left|3-2{y\over x}\right| \right)  \log \left(H_1\over \Mp\right)
-{1\over 2}  \left(3-\left|3-2{y\over x}\right| \right)  \log \left(M\over \Mp\right)>
\nonumber \\ && ~~~~~~~~~~~~~~~~~~~~~~~
-2 +\log \left(2\pi^2 g_1^4\over f^2 \right) + \left(1+\left|3-2{y\over x}\right| \right)\log c_s.
\label{515}
\eea

Let us now consider the intermediate frequency band (\ref{54}). For the ratio  $\om_M/\om_s$ the expression (\ref{511}) is still valid, and for $P_\psi$ (with $c_s=1$) we get the same condition as before, expressed in the limiting case $\om_M=\om_s$. However, if $P_\sg$ is dominant in this band defined by
\beq
{1\over 2} \log \left(M\over \Mp\right) <x- {5\over 6}\log  \left(H_1\over \Mp\right)<{1\over 2} \log \left(M\over \Mp\right)-  \log c_s,
\label{516}
\eeq
we obtain the  new condition  $P_\sg(\om_M)>10^{-2}$, which can be written as 
\beq
x +|3x-2y| -{4\over 3} \log \left(H_1\over \Mp\right) - 2  \log \left(M\over \Mp\right)>
-2 +\log \left(2\pi^2 g_1^4\over f^2 \right).
\label{517}
\eeq

We are left with the lowest frequency branch (\ref{55}), for which we need $\om_M/\om_s$ and $\om_M/\om_{LS}$. For the first ratio we can still apply Eq. (\ref{511}), while for the second ratio, using the previous result (\ref{410}), we obtain:
\beq
{\om_M\over \om_{LS}} ={\om_M\over \om_{s}} {\om_s\over \om_{LS}} 
\simeq {1\over 6} 10^{26} 
\left(M\over \Mp\right)^{-1/2}.
\label{518}
\eeq
If $P_\psi$ is dominant thus 
we have, for this frequency branch, 
\beq
x- {5\over 6}\log  \left(H_1\over \Mp\right)<{1\over 2} \log \left(M\over \Mp\right),
~~~~~~~~~
{1\over 2} \log \left(M\over \Mp\right)<26-\log 6,
\label{519}
\eeq
and the constraint $P_\psi(\om_M)>10^{-2}$gives
\beq
|3x+2y| -{1\over 2} \log \left(H_1\over \Mp\right)
-{3\over 2}  \log \left(M\over \Mp\right)
>-2 +\log (2\pi^2).
\label{520}
\eeq
If $P_\sg$ is dominant the conditions (\ref{519}) still apply, but we have to impose the constraint  $P_\sg(\om_M)>10^{-2}$, which implies:
\bea
&&
|3x-2y| +(n_s-4)x +{17-5n_s\over 6} \log \left(H_1\over \Mp\right)
-{1\over 2}  (n_s-1) \log \left(M\over \Mp\right)
>-2 +\log \left(2\pi^2 g_1^4\over f^2 \right).
\nonumber \\ &&
\label{521}
\eea

We are now in the position of combining all constraints introduced in this Section in order to define, for our class of models,  the allowed region of parameter space compatible with a significant PBH production. The results obviously depend on two crucial physical quantities: the PBH mass $M$, and the sound speed parameter $c_s$.

For our illustrative purpose we shall concentrate on the mass value $M=10^{20}$g $\sim 10^{-13} M_\odot  \sim 10^{24}\Mp$  (but we have checked that the final result is  very little sensitive to a variation of $M$ in the range $10^{18}-10^{22}$ g). Also, since we are interested in the possible overlapping of the PBH-allowed region with the region allowed by all the phenomenological constraints of Sect. \ref{Sec4}, we shall consider numerical values of $c_s$ in the range $c_s\gaq 0.0035$. For smaller
values of $c_s$, in fact, this second allowed region tends to rapidly disappear (as shown in Fig. \ref{f1}). 

%%%%%%%%%%%%%%%%%%%%%%%%%%%%%%%%%%%%%%%%%%%%%%%%%%%%%%%%%%%%%%%%%%%%%%%%%%%
\begin{figure}[t]
\centering
\includegraphics[width=9cm]{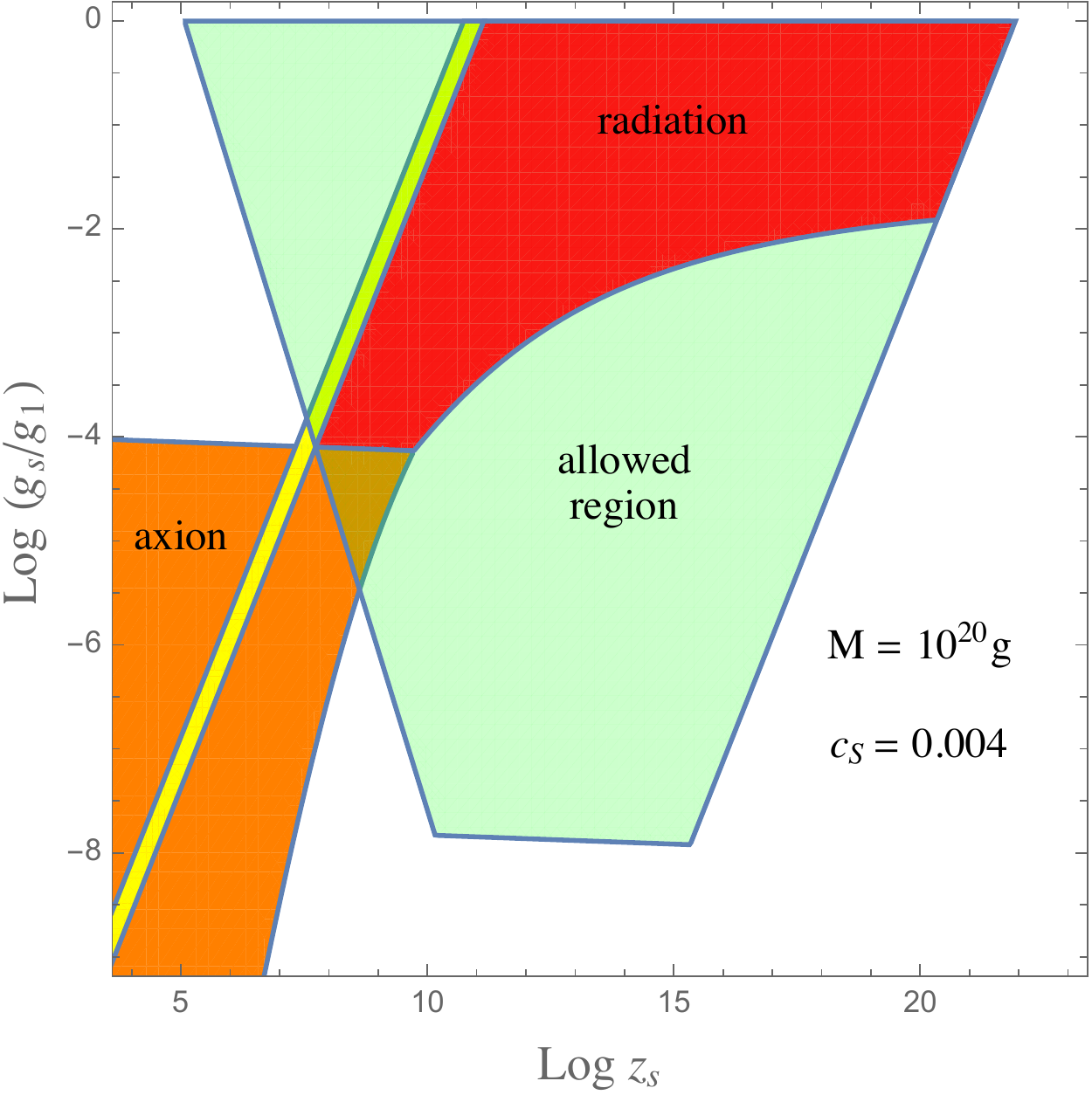}
\caption{The green region allowed by the constraints of Sect. \ref{Sec4} is compared with the region of parameter space compatible with a significant production of PBHs of mass $M=10^{22}$g. This second region is very little sensitive to the mass value for $M$ varying in the range $10^{18}-10^{22}$ g. PBH production may be triggered by scalar perturbation modes leaving the horizon during the string phase and re-entering in the radiation era (red and yellow regions) or during the axion-dominated regime (orange region). As discussed in the text, the overlap requires a fine-tuned range of values for the sound-speed parameter, and the figure refers to the particular value  $c_s=0.004$.} 
\label{f2}
\end{figure}
%%%%%%%%%%%%%%%%%%%%%%%%%%%%%%%%%%%%%%%%%%%%%%%%%%%%%%%%%%%%%%%%%%%%%%%%%%%

The obtained results are illustrated in Fig. \ref{f2} for $c_s=0.004$ and $M= 10^{20}$ g. The region of parameter space compatible with an important PBH production -- namely, satisfying the conditions $P(\om_M)>10^{-2}$ and $P<1$ -- is plotted in orange for the frequency branch (\ref{52}), in red for the frequency branch (\ref{53}), and in yellow for the frequency branch (\ref{54}). As clearly shown by the picture, there is thus a possible interesting contribution from modes amplified by the high-energy string phase, but no contribution at all from modes leaving the horizon during the initial phase of low-energy inflation (the lowest frequency branch (\ref{55})). Also, the largest overlapping with the green region (determined by the phenomenological constraints) is obtained for the string modes re-entering the horizon during the radiation era (red region), while there is only a small overlap for modes re-entering during the initial epoch of axion-dominated evolution (orange region). 

It can be easily checked that the borders of the (red-yellow-orange) regions compatible with PBH production are (continuously and very slightly) shifted towards the left and upwards in the figure as $M$ decreases, and towards the right and downwards as $M$ increases. However, the area of the overlap region keeps practically unchanged for $M$ varying in the relevant range $10^{18}-10^{22}$g.

If we increase the given value of $c_s$ we know, as discussed in Sect. \ref{Sec4}, that the green region tends to slightly increase up to the maximum value corresponding to $c_s=1$ (see Fig. \ref{f1}). However, the extension of the regions compatible with PBH dark matter decreases, and the corresponding overlap with the green region tends to disappear. In particular, we have checked that the overlapping of the orange region (string modes re-entering during the axion era) completely disappears for $c_s \gaq 0.01$. The red region (string modes re-entering during the radiation era) also decreases for growing $c_s$, and for $c_s \gaq 0.013$ it has no more overlap with the relevant sector of parameter space ($y<0$). Similarly, the (very small) yellow region fully disappears from parameter space for $c_s \gaq 0.01$.

Conversely, if $c_s$ is decreased, then the green region rapidly decreases, and  there in no longer any overlap with the orange region for $c_s \laq 0.00325$. The red region also decreases for decreasing $c_s$, and disappears from the allowed range $y<0$ for $c_s \laq 0.00323$. Finally, the yellow region fully disappears for $c_s \laq 0.0032$.

Summarizing the results of this section, we can thus conclude that the considered class of models is  compatible with the possible production of a large amount of PBHs, and in a wide region of parameter space, but this requires a non-negligible amount of ``fine-tuning" for the sound-speed parameter $c_s$, which should be confined, to this purpose, in the rather narrow range of values $0.003 \laq c_s \laq 0.01$.

%%%%%%%%%%%%%%%%%%%%%%%%%%%%%%%%%%%%%%%%%%%%%%%
\section{Conclusion}
\label{Sec6}
\setcounter{equation}{0}
%%%%%%%%%%%%%%%%%%%%%%%%%%%%%%%%%%%%%%%%%%%%%%%

With the preliminary, phenomenological analysis of this paper we have shown that a significant production of dark matter in the form of PBHs, of  mass allowed by present astrophysical constraints, is in principle compatible with a long enough  ``stringy" phase of high-curvature inflation, characterized by an appropriate value of the effective sound-speed parameter.

The underlying dynamical mechanism is very simple, and not new: the spectral amplitude of primordial perturbations $P(k)$ turns out to be enhanced by values $c_s<1$ of the sound-speed parameter appearing with a negative power in the spectrum, thus boosting PBH production. Such a positive effect is limited by the fact that, if $c_s$ is too small, then additional constraints are triggered to avoid the destructive backreaction effects of a too large $P(k)$, and the allowed region of parameter space tends to rapidly disappear. 

As we have stressed, the modification of the spectral amplitude due to $c_s$ in principle occurs for all types of perturbations. For the scenario considered in this paper, however, the most efficient effect is obtained in the case of the primordial axion spectrum, which is much flatter -- hence, more suitable to favor PBH production in the relevant range of masses (and of frequencies) -- than the primordial spectrum of scalar metric perturbations.

For the particular class of models discussed here we have found that a possibly significant PBH production requires $c_s$ to be in the range $0.003 \laq c_s \laq 0.01$. Correspondingly, the required range of the other inflationary parameters is $10^{7} \laq z_s \laq 10^{22}$ for the extension in time of the string phase,  and 
$10^{-6} \laq g_s/g_1  \laq 1$ for the associated growth of the string coupling. Given the above values we find, according to Eq. (\ref{47}), that the final transition scale may vary in the range $10^{-10}\Mp  \laq H_1 \laq 10^{-3}\Mp \sim 10^{15}$ GeV. Of course, the lower limit is not expected to be realistic in the context of a string-theory  model of inflation.

In this paper we have assumed a simple (but motivated \cite{20}) model of background with a step-like variation of the parameter $c_s$, and we have computed the primordial spectra (differently from \cite{19a}) by directly matching the background fluctuations (and not the canonical variable) across the given transition hypersurfaces. However, we are aware of the fact that, in a possibly more realistic scenario (to be studied in a future paper \cite{55}), the high-curvature string phase should include the presence of matter sources \cite{42}, and that the associated sound-speed parameter should be described by an effective function $c_s(\eta)$ possibly varying in time even during the epoch of constant-curvature evolution.

Let us  finally recall that, in the string cosmology scenario that we have considered,  the collapse of primordial inhomogeneities leading to PBH formation can be associated with perturbation modes re-entering the horizon either in the radiation- or in the axion-dominated regime of post-inflationary evolution. In the second case, corresponding to a dust-dominated epoch, it turns out that the spectral constraints 
determining a significant PBH production might be somewhat relaxed \cite{56} (see also \cite{58}) with respect to the one used in this paper ($P \gaq 10^{-2}$). If so, this would produce a corresponding increase in size of the allowed ``orange" region illustrated in Fig. \ref{f2}. The study of this possibility requires however a more detailed discussion, that we have preferred to postpone to a forthcoming paper \cite{57}.

%%%%%%%%%%%%%%%%%%%%%%%%%
% \vskip 1 cm
 
\section*{Acknowledgement}
MG and GM are supported in part by INFN under the program TAsP ({\it Theoretical Astroparticle Physics}). MG is supported in part by the research grant number 2017W4HA7S {\it ``NAT-NET: Neutrino and Astroparticle Theory Network"} under the program PRIN 2017 funded by the Italian Ministero dell'Universit\`a e della Ricerca (MUR). We are grateful to Gabriele Veneziano and Giuseppe Fanizza for a careful reading of the manuscript and many useful  suggestions. Finally, we wish to thank Gabriele Veneziano for an earlier collaboration and important discussions during the preparation of this work.

%%%%%%%%%%%%%%%%%%%%%%%%%%%%
%%%%%%%%%%%%%%%%%%%%%%%%%%%%%%

\end{document}